\documentclass[journal=ancac3,manuscript=article]{achemso}

\usepackage[version=3]{mhchem} 
\usepackage[T1]{fontenc}

\author{E. Rusak}
\affiliation{Institute of Theoretical Solid State Physics, Karlsruhe Institute of Technology, 76131 Karlsruhe, Germany}
\author{J. Straubel}
\affiliation{Institute of Theoretical Solid State Physics, Karlsruhe Institute of Technology, 76131 Karlsruhe, Germany}
\author{P. G\l{}adysz}
\affiliation{Institute of Physics, Faculty of Physics, Astronomy and Informatics, Nicolaus Copernicus University, Grudziadzka 5, 87-100 Torun, Poland}
\author{M. G\"{o}ddel}
\affiliation{Institute of Theoretical Solid State Physics, Karlsruhe Institute of Technology, 76131 Karlsruhe, Germany}
\author{A. K\k{e}dziorski}
\affiliation{Institute of Physics, Faculty of Physics, Astronomy and Informatics, Nicolaus Copernicus University, Grudziadzka 5, 87-100 Torun, Poland}
\author{M. K\"{u}hn}
\affiliation{Institute of Physical Chemistry, Karlsruhe Institute of Technology, 76131 Karlsruhe, Germany}
\author{F. Weigend}
\affiliation{Institute of Nanotechnology, Karlsruhe Institute of Technology, 76021 Karlsruhe, Germany}
\author{C. Rockstuhl}
\affiliation{Institute of Theoretical Solid State Physics, Karlsruhe Institute of Technology, 76131 Karlsruhe, Germany}
\alsoaffiliation{Institute of Nanotechnology, Karlsruhe Institute of Technology, 76021 Karlsruhe, Germany}
\author{K. S\l{}owik}
\affiliation{Institute of Physics, Faculty of Physics, Astronomy and Informatics, Nicolaus Copernicus University, Grudziadzka 5, 87-100 Torun, Poland}

\email{karolina@fizyka.umk.pl}
\phone{+48 56 611 3329}

\title[hot]
  {Enhancement of and Interference among Higher
Order Multipole Transitions in Molecules near
Plasmonic Nanoantenna
}


\begin{document}







\begin{abstract}
Spontaneous emission of quantum emitters can be modified by engineering their optical environment. This allows a resonant nanoantenna to significantly modify the radiative properties of a quantum emitter. In this article, we go beyond the common electric dipole approximation for the molecular electronic transition and take light-matter coupling through higher order multipoles into account. We investigate, by means of theory and numerical simulations, a strong enhancement of the magnetic dipole and electric quadrupole emission channels of a molecule adjacent to a plasmonic patch nanoantenna. While this on its own had been considered, the assumption in prior work usually has been that each molecular transition is dominated only by one of those multipolar emission channels. This leads naturally to the notion of discussing the modified emission in terms of a modified local density of states defined for each specific multipolar transition. In reality, this restricts the applicability of the approach, since specific molecular transitions occur via multiple multipolar pathways that have to be considered all at once. Here, we introduce a framework to study interference effects between higher order transitions in molecules by (a) a rigorous quantum-chemical calculation of their multipolar moments and (b) by a consecutive investigation of the transition rate upon coupling to an arbitrarily shaped nanoantenna. Based on that formalism we predict interference effects between these transition channels. This allows for a strong suppression of radiation by exploiting destructive interference. Our work suggests that placing a suitably chosen molecule at a well defined position and at a well defined orientation relative to a nanoantenna can fully suppress the transition probability.  \end{abstract}

\section{Introduction} 
The spontaneous emission rate of quantum emitters is not an intrinsic property: it depends on their optical environment. 
In recent years, optical nanoantennas have been suggested to modify the radiative properties of point sources. \cite{Pelton2015}. 
Motivated by their radiofrequency counterparts, optical nanoantennas, with sizes in the order of hundreds of nanometers, are antennas for light. An optical nanoantenna is commonly defined as a device designed to efficiently convert free-propagating optical radiation to localized energy, and vice versa. Localized energy or a strongly confined field is present if an electromagnetic field is concentrated in a tiny spatial domain when compared to the free space wavelength of light. In plasmonics, metallic nanoantennas are used to confine electromagnetic fields to volumes smaller than the diffraction limit. This happens as the conduction electrons of the plasmonic nanoantenna can be driven by the incident electric field in a resonant collective oscillation known as a surface plasmon polariton.

The huge plasmonic field enhancement and related spatial energy confinement leads to a boost of the energy exchange rate between a quantum emitter and the electromagnetic field. Resulting shortened interaction timescales are, in particular, the reason for a suppression of spontaneous emission lifetimes through the Purcell effect by multiple orders of magnitude \cite{Hoang2015}. For quantum emitters located only a few nanometers from the nanoantenna surface a quenching effect was reported \cite{Anger2006}. We emphasize that it does not imply suppressed spontaneous emission lifetimes: an important reason for quenching is that the spontaneously emitted photons are rather absorbed by the nanoantenna than radiated into the far field \cite{Bohren}. Still, the transition occurs at a Purcell-enhanced rate. Here we add an additional layer of complexity to the control of spontaneous emission rates, i.e. either its enhancement or suppression, through tailored interference of different light-matter interaction channels. 

Usually, only the electric dipole contribution to a quantum-mechanical state transition of a quantum emitter is considered. This is often justified by the negligible spatial variation of the electric field over the size of the quantum emitter \cite{Scully}, although recently, studies have been made of spatially extended emitters adjacent to plasmonic nanoantennas  \cite{Neuman2018}.
In presence of a nanoantenna, the electric field is localized into nanometric spatial domains, providing high intensities and spatial modulations at the length scale of a molecule \cite{Gramotnev2010}.  Thus, higher order multipolar contributions to light-matter coupling may become relevant. Until now, the enhancement by a nanoantenna of magnetic dipole emission was studied both theoretically and experimentally \cite{Karaveli2010,Taminiau2012,Hein2013,Aigouy2014,Kasperczyk2015}. 
Large enhancements of electric quadrupole transitions were also predicted \cite{Kern2012,Filter2012,Yannopapas2015}. Transitions driven with several multipolar mechanisms have been observed in semiconductor quantum dots \cite{Tighineanu2014}. 

In these prior works, the considered higher order multipolar transitions were usually assumed to be either purely electric or magnetic dipolar, or purely electric quadrupolar. However, depending on their symmetry properties dictated by their geometry, quantum emitters can have transitions with contributions from different multipoles at once. In principle, each of these multipoles can be enhanced near a nanostructure, as predicted in Ref.~\citenum{Rivera2016} and demonstrated experimentally in Ref.~\citenum{Li2018}. If these contributions correspond to a transition between the same pair of eigenstates, interference effects between the emission of a photon via different transition channels are expected \cite{Craig}. In Ref.~\citenum{Tighineanu2014}, the ability of semiconductor quantum dots to probe electric and magnetic fields simultaneously was shown, stating interference between higher order decay channels of the quantum dots. 

In this work, we study the interference effects between different multipole transition channels by coupling a molecule to a plasmonic nanoantenna in a conceptually and operationally simple approach based on Fermi's golden rule. We consider a patch nanoantenna of a geometry rich enough to allow for different effects to be discussed, from selective enhancement of different multipolar contributions, to constructive or destructive interference thereof. We demonstrate the tunability of such nanoantenna with respect to magnetic-dipole and electric-quadrupole enhancement in a range of wavelengths from mid-optical to near-infrared. Next, the geometry is tuned to a specific transition of a specific molecule, characterized by a set of multipolar moments, that have been calculated by means of quantum-chemical methods. Depending on the molecular orientation, control of transition rate enhancement in various channels is demonstrated. Consequently, we show that in deliberately chosen systems, i.e. combinations of molecules and optical nanoantennas, it must be possible to suppress the emission comparable to what has been suggested long time ago while using spatially extended photonic crystal structures with a well defined optical band gap \cite{Noda2007}. 

\subsection{Different multipolar contributions to transition rates}
Our approach is based on Fermi's golden rule, which accounts for the molecular transition rate depending on the nanoantenna-scattered electromagnetic field properties and of molecular characteristics. 
A theoretical prediction for the transition rate $\Gamma$ from an
excited state $|\mathrm{i}\rangle$ with energy $\hbar \omega_{\mathrm{i}}$ to a final state $|\mathrm{f}\rangle$ of energy $\hbar\omega_\mathrm{f}$ is given by \cite{Loudon}
\begin{equation}\label{eq:golden_rule}
\Gamma = \frac{2\pi}{\hbar^2}\left|\langle\mathrm{f}|\mathcal{V}|\mathrm{i}\rangle\right|^2 \rho\left(\omega_\mathrm{i}-\omega_\mathrm{f}\right),
\end{equation}
with the reduced Planck constant $\hbar$. The density of states $\rho\left(\omega\right)$ needs to be taken at the quantum emitters transition frequency \mbox{$\omega_\mathrm{i}-\omega_\mathrm{f}$}. In this work, we make a step beyond the common electric dipole approximation and take higher order contributions into account. Thus, the interaction Hamiltonian $\mathcal{V}$ up to the electric-quadrupolar order is studied \cite{Barron1973}
\begin{equation} \label{eq:hamiltonian}
\mathcal{V} = -\underbrace{\mathbf{p} \cdot \mathbf{E}\left(\mathbf{r}_0\right)}_{\mathcal{V}_\mathrm{ED}} - \underbrace{\mathbf{m} \cdot \mathbf{B}\left(\mathbf{r}_0\right)}_{\mathcal{V}_\mathrm{MD}} - \underbrace{\left[\mathbf{Q}\nabla\right] \cdot \mathbf{E}\left(\mathbf{r}_0\right)}_{\mathcal{V}_\mathrm{EQ}},
\end{equation}
where $\mathbf{r}_0$ is the location of the molecule, and the electric dipole ED, magnetic dipole MD and electric quadrupole EQ contributions are included\footnote{Please note that in the above Hamiltonian the elements of the quadrupole moment operator are defined as $Q_{kl} = \frac{e}{2} r_k r_l $, but they can be replaced by the traceless form $Q_{kl}=\frac{e}{2}( r_k r_l-\frac{1}{3}\delta_{kl}r^2)$ in source-free regions, where $\nabla\cdot \mathbf{D} =0$.}. 
The Hamiltonian in Eq.~\ref{eq:hamiltonian} includes the electric $\mathbf{E}$ and magnetic $\mathbf{B}$ fields induced by a combination of considered multipolar sources, scattered by the nanoantenna and evaluated at the molecular position. The fields are calculated classically (for technical details please see \textit{Methods: field characteristics} and \textit{Supporting Information}). The transition moments in Eq.~\ref{eq:hamiltonian}, i.e. the matrix elements of the electric dipole $\mathbf{p}$, magnetic dipole $\mathbf{m}$, and electric quadrupole $\mathbf{Q}$, are calculated between the initial and final states of the molecule with time-dependent density functional theory (TDDFT), as described  in \textit{Methods: Molecular characteristics}. The square bracket in Eq.~\ref{eq:hamiltonian} contains a matrix multiplication of the electric quadrupole moment tensor and the column vector \mbox{$\nabla = \left( \frac{\partial}{\partial x},\frac{\partial}{\partial y},\frac{\partial}{\partial z}\right)^T$}, while $T$ denotes operation of transposition.

To describe spontaneous emission, the interaction Hamiltonian (\ref{eq:hamiltonian}) should account for single-molecule-to-single-photon coupling\cite{Scully}, i.e. the electromagnetic fields around the nanoantenna should be normalized to values corresponding to single-photon excitations. Normalization to single-photon fields assures that the resulting transition rate $\Gamma$ scales with the square of transition moments as expected. 

Proper normalization factor can be found based on free-space emission rates for different multipoles and the corresponding Purcell enhancement factors 
$F^{\mathcal{M},\phi}_\mathrm{tot}(\mathbf{r}_0,\omega)=P^{\mathcal{M,\phi}}_\mathrm{na}(\mathbf{r}_0,\omega)/P^{\mathcal{M}}_0(\omega)$. Purcell factors are classically calculated for each type of emitter, by evaluation of powers of the emitter in the presence of the nanoantenna $P^{\mathcal{M,\phi}}_\mathrm{na}(\mathbf{r}_0,\omega)$ and in free space $P^{\mathcal{M}}_0(\omega)$. Here $\mathcal{M} \in \{\mathrm{ED},\mathrm{MD},\mathrm{EQ}\}$ denotes the type of emitter, and $\phi$ indicates that the result depends on the orientation of the emitter with respect to the nanoantenna. 
The total Purcell factor describes how much the extracted power from an oscillating dipole or quadrupole is enhanced in the presence of a photonic nanostructure when compared to free space. 
Identification of the emitted powers rescaled by single-photon energy $P^{\mathcal{M},\phi}_\mathrm{na}/\hbar \omega$ ($P^{\mathcal{M}}_0/\hbar \omega$) with the transition rates $\Gamma^{\mathcal{M},\phi}$ ($\Gamma_0^{\mathcal{M}}$) of a quantum emitter (e.g. a molecule) allows us to write 
\begin{equation}\label{eq:calibration_ed}
\underbrace{\frac{2\pi}{\hbar^2} \left|\langle\mathrm{f}|\mathcal{V_\mathrm{ED}}|\mathrm{i}\rangle\right|^2 \rho}_{\Gamma^{\mathrm{ED},\phi}} = F^{\mathrm{ED},\phi}_\mathrm{tot} \underbrace{\frac{\omega^3 |\mathbf{p}|^2}{3\pi\epsilon_0\hbar c^3}}_{\Gamma^\mathrm{ED}_0},
\end{equation}
where we have used Eq.~\ref{eq:golden_rule} and the definition of the total Purcell enhancement factor. Above, $\epsilon_0$ stands for the vacuum electric permittivity, and the expression for $\Gamma^\mathrm{ED}_0$ on the right-hand side is given by the Weisskopf-Wigner formula \cite{Scully}. Similarly, for magnetic-dipole and electric-quadrupole sources we have 
\begin{eqnarray}\label{eq:calibration_mdeq}
\underbrace{\frac{2\pi}{\hbar^2} \left|\langle\mathrm{f}|\mathcal{V_\mathrm{MD}}|\mathrm{i}\rangle\right|^2 \rho}_{\Gamma^{\mathrm{MD},\phi}} &=& F^{\mathrm{MD},\phi}_\mathrm{tot} \underbrace{\frac{\omega^3 |\mathbf{m}|^2}{3\pi\epsilon_0\hbar c^5}}_{\Gamma^\mathrm{MD}_0},\\
\underbrace{\frac{2\pi}{\hbar^2} \left|\langle\mathrm{f}|\mathcal{V_\mathrm{EQ}}|\mathrm{i}\rangle\right|^2 \rho}_{\Gamma^{\mathrm{EQ},\phi}} &=& F^{\mathrm{EQ},\phi}_\mathrm{tot} \underbrace{\frac{\omega^5 \sum_\mathrm{i,j}|\mathbf{Q}_\mathrm{ij}|^2}{10\pi\epsilon_0\hbar c^5}}_{\Gamma^\mathrm{EQ}_0},\label{eq:calibration_eq}
\end{eqnarray}
where on the right-hand side we have used the free-space MD and EQ transition rates \cite{Craig}.
Based on this equality we can jointly normalize fields in $\mathcal{V}_\mathcal{M}$ and density of electromagnetic states $\rho$. Since any kind of multipole can in general be a source of both electric and magnetic fields, to be scattered by the nanoantenna, we normalize each field from a given source by the same factor, preserving phase relations. Obviously, with the proposed scheme we retrieve expected free-space transition rates. 
Based on these normalized single-photon fields, we can also consider subsequently an arbitrary superposition thereof that requires a weighting of each term according to the multipolar moments of the transitions of a specific molecule. These multipolar moment elements are obtained from quantum-chemical simulations (see below). Please note, when reconstructing the single-photon fields that are considered in Eq.~\ref{eq:hamiltonian}, the contribution of each multipolar emitter to the electric and magnetic field and the electric field gradient is taken into account. We use the total field, i.e. the one emitted by the actual multipolar source and the secondary field scattered by the antenna. The procedure to extract the fields at the point of interest is described in the \textit{Supporting Information}.

Before we continue to discuss results, we make a comment on quantifying influence of the photonic environment, i.e. the nanoantenna. 
Canonically, the influence of nanoantennas or cavities in general on spontaneous emission of quantum systems is expressed in terms of modified density of states $\rho\left(\omega\right)$ \cite{novotny_book}.
Once a source is defined, it accounts for field enhancement at the source's location and spectral dependencies.
We stress that in the method described in this work, in contrast to the usual approach, the same influence is taken into account through the modification of field profiles $\mathbf{E}\left(\mathbf{r}\right)$ and $\mathbf{B}\left(\mathbf{r}\right)$: it is the field that is enhanced and reshaped, rather then density of states.
This allows us to assume that the density of states is common for all transition mechanisms.
Naturally, it makes only sense to consider the impact of the photonic environment in only one of the two terms, $\mathcal{V}$ and $\rho$, that appear in Eq.~\ref{eq:golden_rule}. Here, in the presence of multiple multipolar transitions channels, we found it more convenient to account for  the effect of the proximity of nanoantenna via the modification of the field within the matrix elements appearing in the respective expressions. This allows us to account for potentially significant interference effects between different transition mechanisms that contribute to the Hamiltonian (\ref{eq:hamiltonian}). 

\section{Results and discussion}
In this section we introduce an exemplary highly symmetric molecule and provide its multipolar characteristics determined by TDDFT. These characteristics are given in terms of elements of the multipolar transition moments of suitably chosen electronic transitions occuring in the optical regime.
Next, we provide details of a nanoantenna geometry suited to resonantly enhance two higher-order multipolar channels: the magnetic dipolar and the electric quadrupolar ones. We discuss the spectral positions of the nanoantenna resonances due to illumination with the corresponding types of sources. Finally, we focus on a specific nanoantenna whose dimensions were tuned to the specific molecular transition to demonstrate various functionalities that can be achieved, including enhancement of predefined transition channels (here MD or EQ) or their cooperative suppression by means of interference between different transition amplitudes. 

\subsection{Description of molecule}
\begin{figure}
 \includegraphics[width=8.6cm]{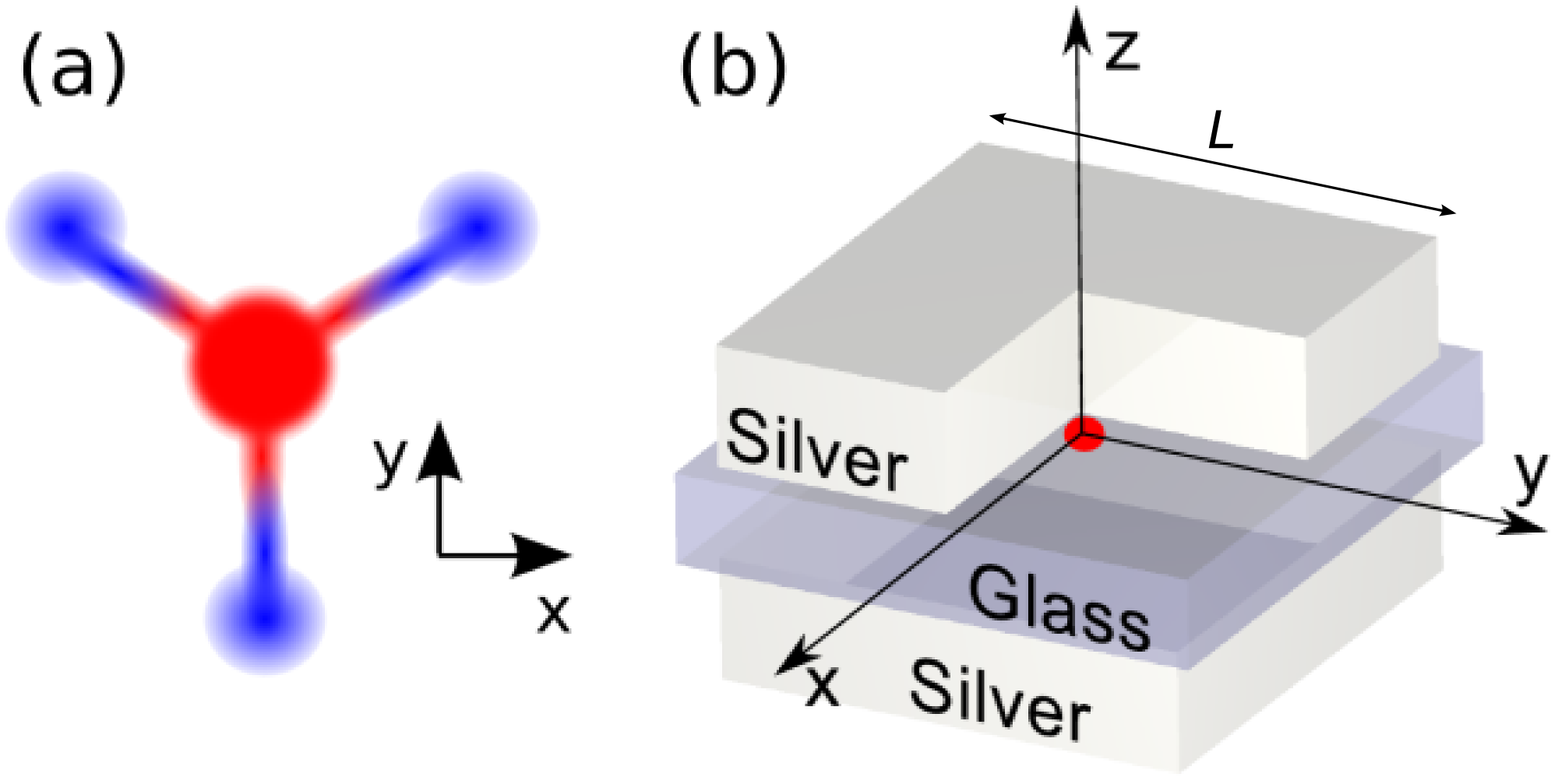}
  \caption{(a) Geometry of an OsO$_3$ molecule. The osmium atom is sketched as a red, and oxygen atoms as blue balls. All atoms are positioned in the $xy$ plane.
  (b) Geometry of the considered nanoantenna: two cuboid silver patches of size $L\times L\times 50$ nm, separated with a $30$ nm thick dielectric spacer with a side length $L + 20$ nm. The red dot indicates the molecule placed in the center of the nanoantenna. Please note the upper cuboidal patch is identical to the lower one, while the cut in the above figure serves only visualization purposes, so that the location of the molecule can be seen.}
  \label{fig:molecule_nanoantenna}
\end{figure}

Let us first focus on an exemplary system that matches our requirements, i.e. the TDDFT-based model of an OsO$_3$ molecule, whose geometry is sketched in Fig.~\ref{fig:molecule_nanoantenna}(a). The molecule is of the $D_{3h}$ symmetry that defines the selection rules for optically-driven state transitions \cite{Gelessus1995}. According to these selection rules, only certain matrix elements of multipolar moment operators, calculated for the states between which the transition occurs, can be nonzero \cite{Hamermesh, Cotton}. We have obtained these values, as it is mentioned above, using the TDDFT method \cite{Furche2014} (for details please see \textit{Methods: Molecular characteristics}; please also note there is a comment on robustness of these results with respect to applying different exchange-corrlation functionals within DFT or small geometry changes). 

The first (degenerate) pair of the excited states obtained within TDDFT method is roughly $1.5$ eV above the ground state, but the ED, MD and EQ transition moments are zero. This is because these transitions result from spin-forbidden singlet-triplet excitations from the non-degenerate HOMO to the twofold degenerate LUMO of $e^{\prime\prime}$ symmetry.
The next degenerate pair of electronic states of the molecule was found at $2.24$ eV above the ground state, resulting from the corresponding (spin-allowed) singlet excitations. This corresponds to the free-space excitation wavelength of $553$ nm, and is well isolated from the next electronic transition at $2.98$ eV ($416$ nm).
The two degenerate states at $553$ nm, which we will denote with indices $\mathrm{j}=1, 2$, form a basis of a two-dimensional irreducible representation $E''$ of $D_{3h}$, while the ground state $0$ is fully symmetric ($A^\prime_1$). In theory, the allowed transitions between the ground and the excited states, i.e. those with nonzero transition matrix elements, are those coupled by the $m_x$ and $m_y$ components of the magnetic dipole moment, and $Q_{yz}$ and $Q_{xz}$ components of the electric quadrupolar moment\cite{Hamermesh, Cotton}. 
Indeed, we find in the TDDFT simulations both transitions between the ground state $0$ and the excited states $\mathrm{j}=1,2$ introduced above, which we will refer to as \textit{transition 1} or \textit{transition 2}, to be electric-dipole forbidden. The components of their electric dipole moments $\mathbf{p}^{0\mathrm{j}}$, that would enter the Hamiltonian in Eq.~\ref{eq:hamiltonian}, are zeros. 

The only nonzero components of the magnetic dipole $\mathbf{m}^{0\mathrm{j}}$ and electric quadrupole $\mathbf{Q}^{0\mathrm{j}}$ moments are $m_y^{01} = im$, and  $Q_{xz}^{01} =Q$ 
for transition $1$, and $m_x^{02} = im$ and $Q_{yz}^{02} = -Q$ for transition $2$. Here, $i$ stands for the imaginary unit, $m= 0.420$ a.u. $\approx 7.79 \times 10^{-24}$ J/T and $Q = 0.16$ a.u. $\approx 7.2\times 10^{-41}$Cm$^2$. 
According to Eqs.~(\ref{eq:calibration_mdeq}), resulting free space spontaneous emission rates for the magnetic dipole and electric quadrupole channel read respectively $\Gamma^\mathrm{MD}_0\approx 112.6$ Hz and $\Gamma^\mathrm{EQ}_0\approx 0.067$ Hz.
This indicates that the quadrupolar channel is rather weak and the magnetic dipolar character dominates.

In the next subsection we describe a nanoantenna designed to match the character of this particular molecule: it sustains a resonant optical response to a magnetic dipolar source, and a significantly stronger one with respect to an electric quadrupolar illumination. In this way the nanoantenna is aimed to restore balance between strengths of the two transition mechanisms.

Moreover, in the following sections we will consider the molecule to be positioned in a dielectric matrix. Atomic-scaled inhomogeneities in the surroundings might in principle trigger breaking of selection rules, giving rise to nonzero, but small, electric dipole moments. Furthermore, the fact (not taken into account here) that the symmetry of the molecule being in one of its emitting excited states may be lower than $D_{3h}$ due to the Jahn-Teller effect \cite{Jahn}, would change the selection rules of the considered transitions. The inclusion of the spin-orbit interaction into the theoretical model would slightly modify the physical character of the excited electronic states of OsO$_3$ molecule and, as a consequence, the selection rules and the transition moments.
Finally, transitions from singlet states $1$ and $2$ to the lower pair of corresponding triplet excited states are not taken into account in our analysis. Radiative transitions between these states are spin-forbidden and thus probably not very relevant. However, intersystem crossing into the lower triplet states, i.e. a nonradiative relaxation, might quench the emission from the states under interest. 
For all these reasons it is important that the nanoantenna does not support electric dipole sources, limiting their potential influence. 

\subsection{Tuning nanoantenna for selected multipolar sources: Purcell enhancement}
The nanoantenna that matches all above requirements consists of two silver patches and a dielectric spacer in between, similar to the one proposed in Ref.~\citenum{Feng2011} for a gold nanoantenna where its magnetic response was investigated. A schematic of the structure is shown in Fig.~\ref{fig:molecule_nanoantenna}(b). This specific geometry was chosen, because besides a strong enhancement of emission by a  magnetic dipole source \cite{Feng2011}, it also provides significant electric field modulations relevant for enhancement of the quadrupolar channel, as we show in the following. 
Since the considered transition wavelength is in the optical regime, silver was chosen rather than gold due to strong absorption in gold at optical frequencies.
Silver was modelled using the experimental data from Ref.~\citenum{JohnsonChristy}. The size $L$ of the quadratic patches is varied in the simulations, while their thickness is set to $50$ nm.
To avoid unphysically sharp edges, the silver patches are modelled as rounded with a radius of curvature of $5$ nm. The gap between the patches is $30$ nm wide, and is filled with a symmetrically positioned quadratic dielectric spacer of length $L + 20$ nm. The permittivity of the dielectric is chosen as $\epsilon = 2.25$. 

In this subsection we study the response of the described nanoantenna to different multipolar sources, i.e. an electric dipole, magnetic dipole and electric quadrupole of basic orientations with respect to the nanoantenna geometry (for details of implementation please see \textit{Supporting Information}). These basic orientations are parallel to the $x$ or $z$ axis of the coordinate frame from Fig.~\ref{fig:molecule_nanoantenna} (in case of dipoles or quadrupole with only diagonal elements), and in the $yz$ or $xy$ plane (in case of off-diagonal quadrupoles). 

\begin{figure}
 \includegraphics[width=\textwidth]{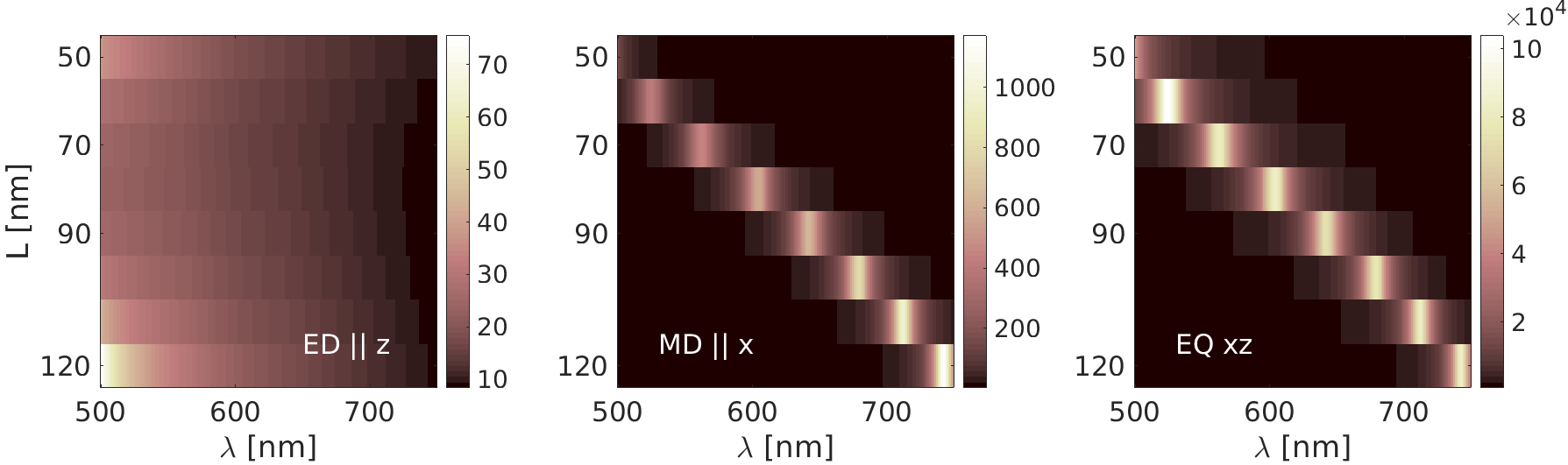}
  \caption{Total, i.e. radiative and absorptive, decay rate enhancement factors $F_\mathrm{tot}$ for different, optimally oriented multipolar sources: 
  (a) electric dipole parallel to the $z$ axis, 
  (b) magnetic dipole parallel to the $x$ axis, 
  (c) off-diagonal electric quadrupole in the $xz$ plane.
  The enhancement is presented in function of free-space wavelength $\lambda$
  of the source and for some discrete nanoantenna patch sizes $L$.}
  \label{fig:resonances}
\end{figure}

The response of the nanoantenna to different sources is studied in terms of the radiative decay rate enhancement $F^{\mathcal{M},\phi}_\mathrm{rad}(\mathbf{r}_0,\omega)=P^{\mathcal{M,\phi}}_\mathrm{na, rad}(\mathbf{r}_0,\omega)/P^{\mathcal{M}}_0(\omega)$
where $P^{\mathcal{M,\phi}}_\mathrm{na, rad}(\mathbf{r}_0,\omega)$ is the time-averaged power radiated by an $\mathcal{M}$-type emitter oscillating at frequency $\omega$ and located at position $\mathbf{r}_0$ near the nanoantenna, and $P^{\mathcal{M}}_0(\omega)$ is the time-averaged power radiated by the emitter in free space. 
In Fig.~\ref{fig:resonances}, a response to different multipolar sources positioned in the center of the nanoantenna is plotted in function of 
frequency $\omega$ and the size $L$ of the nanoantenna patches. Out of the basic orientations of each type of source, in Fig.~\ref{fig:resonances} we only show results obtained for the one that provides the strongest response. Complete characteristics of radiative and total Purcell factors for all sources in different orientations can be found in \textit{Supporting Information}. Please note, only some discrete patch sizes $L$ are considered as evident from the picture.

The nanoantenna sustains a resonant response when illuminated with a magnetic dipole or an off-diagonal electric
quadrupole source, and the resonant wavelength is tunable with $L$. The enhancement factor is significant and reaches over two orders of magnitude in the magnetic, and over four orders of magnitude in the electric quadrupolar case. Resonances due to these two types of sources spectrally overlap and the quadrupolar one dominates by two orders of magnitude - both properties suited to match the requirements of the OsO$_3$ transition indicated in the previous subsection. 
Contrary, the electric dipole and the diagonal electric quadrupole sources do not show resonances in the considered parameter range. This is actually an advantage, since the possible influence of the electric dipole channel that might be unlocked due to local inhomogeneities in the molecular surroundings, is suppressed. 

In this subsection we have described a nanoantenna geometry suited to match all requirements defined in the previous section and based on the properties of the chosen molecule. Power of isolated types of sources, solely magnetic dipole and solely electric quadrupole, can be resonantly enhanced by several orders of magnitude by the proposed nanoantenna. In the following subsection we proceed to analyze a response to a source that combines the magnetic dipole and electric quadrupole components of orientations and strengths corresponding to the OsO$_3$ transitions indicated above. 

\subsection{Interference of multiple transition channels due to combined sources: transition rates }
Based on the results in Fig.~\ref{fig:resonances}, we now fix the nanoantenna patch size at $L = 75$ nm to localize its magnetic dipolar and electric quadrupolar resonances around the OsO$_3$ transition 
wavelength of $552.8$ nm, corresponding to frequency of $2\pi\times 542.4$ ps$^{-1}$. 

Transition rates $\Gamma_\mathrm{0j}$ for each transition $\mathrm{j}=1$ and $\mathrm{j}=2$ were obtained for different orientations of a source, positioned in the center of the nanoantenna, in terms of Fermi's golden rule [Eq.~\ref{eq:golden_rule}], with electric and magnetic fields calculated as described in detail in \textit{Methods:Field characteristics} and \textit{Supporting Information}, and normalized to single-photon strengths according to the method described above in section \textit{Different multipolar contributions to transition rates}. 

The interaction Hamiltonian in Eq.~\ref{eq:hamiltonian} consists of multiple contributions that can interfere. The relative impact of interference effects on the transition rate is expressed through the ratio
\begin{equation}
\mathcal{R} =\frac{\Gamma - \left(\Gamma^\mathrm{ED} + \Gamma^\mathrm{MD} + \Gamma^\mathrm{EQ}\right)}{\Gamma^\mathrm{ED} + \Gamma^\mathrm{MD} + \Gamma^\mathrm{EQ}}.
\end{equation}A positive (negative) value of $\mathcal{R}$ indicates a constructive (destructive) interference, $\mathcal{R}=-1$ denotes a completely destructive process, $\mathcal{R}=0$ means interference effects are absent, and $\mathcal{R}=1$ is a fully constructive effect.

\begin{figure}
 \includegraphics[width=\textwidth]{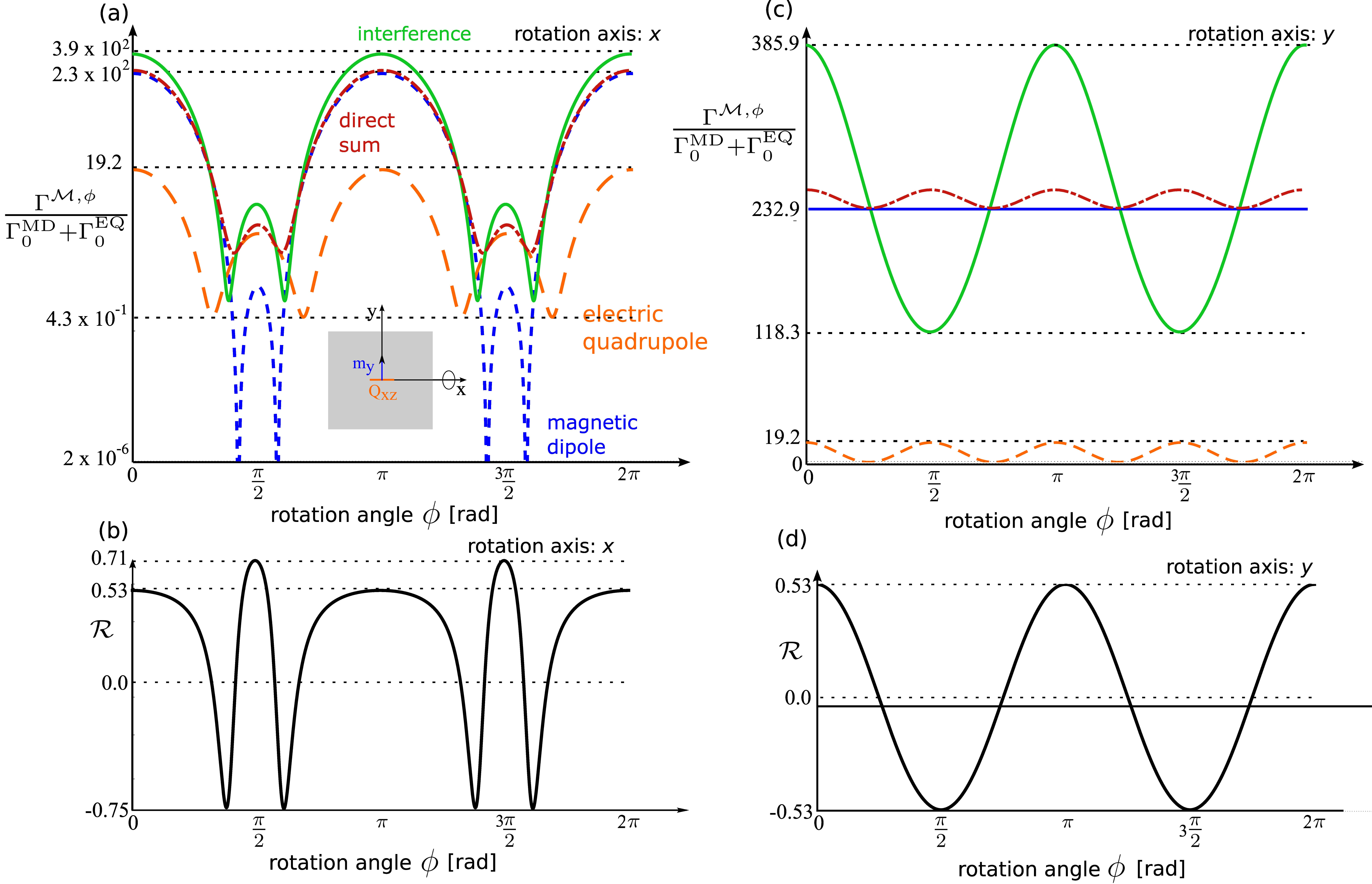}
  \caption{Transition characteristics due to different multipolar mechanisms of a model OsO$_3$ molecule positioned in the center of the nanoantenna, in function of rotation angle with respect to $x$ (b \& c) or $y$ axis (c \& d). Original orientation of multipolar moments of transition $1$ is depicted in the inset of panel (a). 
  (a) Transition rates between the ground state $0$ and excited state $\mathrm{j}=1$. The magnetic dipole $\Gamma^{\mathrm{MD},\phi}$ and electric quadrupole $\Gamma^{\mathrm{EQ},\phi}$  contributions are shown with dotted blue and dashed orange lines, respectively. The total transition rate $\Gamma$ (green solid line) includes the interference between the two contributing mechanisms, and may differ from their direct sum $\Gamma^{\mathrm{MD},\phi}+\Gamma^{\mathrm{EQ},\phi}$ (red dash-dotted line). 
  (b) Figure of merit $\mathcal{R}$ indicating the respective role of the interference terms for the full transition rate shown in (a).
  (c,d) As in (a,b), but for rotations of the molecule around the $y$ axis of the coordinate frame. }
  \label{fig:stable}
\end{figure}

Figure~\ref{fig:stable} shows the transition rates [panels (a \& c)] and impact of interference effects [panels (b \& d)] evaluated for a source, representing an OsO$_3$ molecule, positioned in the center of the nanoantenna, but rotated around axis $x$ [panels (a \& b)] or $y$ [panels (c \& d)]. 
The initial orientation of the molecule (more precisely: transition elements of its multipolar moments $\mathbf{m}$ and $\mathbf{Q}$), shown in the inset of panel (a), is such that the magnetic dipole moment of transition $1$ is parallel to the nanoantenna patches and to the $y$ axis of the coordinate system. This orientation is most favorable in the context of enhancement, as demonstrated for selected rotation axes in Fig.~\ref{fig:stable}. In panel (a) of Fig.~\ref{fig:stable} we demonstrate the transition rates as functions of a rotation angle $\phi\in(0,2\pi)$ around the $x$ axis of the coordinates frame, normalized to the free-space transition rate $\Gamma_0^\mathrm{MD}+\Gamma_0^\mathrm{EQ}$ to demonstrate the respective influence of the nanoantenna. In panel (a) we have used logarithmic scale for better visibility of effects at different scales. The magnetic dipole $\Gamma^{\mathrm{MD},\phi}_{01}$ (dotted blue lines) and electric quadrupole $\Gamma^{\mathrm{EQ},\phi}_{01}$(dashed orange lines) mechanisms contribute to the total transition rate $\Gamma_{01}$ (green solid lines). For a comparison, a direct incoherent sum $\Gamma^{\mathrm{MD},\phi}_{01}+\Gamma^{\mathrm{EQ},\phi}_{01}$ is shown in red dash-dotted line. We emphasize that the latter is not the complete characteristic of the transition, since the direct sum does not include interference effects. 

Our first observation is that in the presence of the nanoantenna, the transition rates through the initially weak channels, commonly considered as "forbidden", reach the regime of several tens of kHz.
In free-space, the transition rates for given moments $\mathbf{m}_{01}$ and $\mathbf{Q}_{01}$ are \mbox{$\Gamma_0^\mathrm{MD}\approx 112.6$ Hz} and \mbox{$\Gamma_0^\mathrm{EQ}\approx 0.067$ Hz}. With the discussed nanoantenna, they are enhanced to \mbox{$\Gamma_{01}^{\mathrm{MD},\phi=0} \approx 26.2$ kHz} in the magnetic-dipole and \mbox{$\Gamma_{01}^{\mathrm{EQ},\phi=0} \approx 2.17$ kHz} in the electric-quadrupole channel. Since we are not able to rotate independently the magnetic dipole and electric quadrupole moment of the transition, for all orientations the response is dominated by the magnetic channel, while the electric quadrupolar one contributes mainly through the interference term. 
As evident from Fig.~\ref{fig:stable}(a,c), for most orientations of the molecule rotated around the $x$ axis, the two channels constructively interfere leading to a further increased full transition rate of up to $\Gamma_{01}^\mathrm{max}\approx 43.5$ kHz, with $\mathcal{R}=53.1\%$ enhancement due to interference. For the most favourable molecular orientations, i.e. with the magnetic dipole moment parallel to the nanoantenna patch edges, the total transition rate exceeds the free-space one by $385.9$ times. Strongest constructive interference reaches $70.6\%$, but happens at orientations for which the transition rate is small and the constructive effect is less influencial. However, even more interesting is the perspective of destructive interference, leading to transition rate suppression. In particular, around \mbox{$\phi\approx (0.5\pm 0.13)\pi$ rad} and \mbox{$(1.5\pm 0.08)\pi$ rad} the phase difference between the magnetic dipolar and the electric quadrupolar response provides destructive interference with $\mathcal{R}=-74.8\%$ and transition rate minima of $\Gamma_{01}^\mathrm{min} \approx 73.23$ Hz, i.e. $64.5\%$ of the free space rate due to magnetic dipole plus electric quadrupole mechanisms\footnote{Please note that in free space no interference between different multipolar mechanisms can occur, since a given type of emitter only gives rise to its characteristic field distribution decoupled from other multipolar sources at the same location.}. Both the free space rate and the nanoantenna-suppressed value are very small with respect to typical electric dipole ones and correspond to large molecular lifetimes of the order of a hundredth of a second. The reason why the interference is not complete and transition is not fully suppressed is the non-optimal phase difference between the magnetic and electric fields at the molecular location, which could be improved through refined engineering. Please note that analysis of other sources of decoherence and dephasing, e.g. due to coupling to phononic bath, is beyond the scope of this work. 

As the molecule rotates around the $y$ [Fig.~\ref{fig:stable}(c,d)] and $z$ (not shown) axis of the coordinate frame, the magnetic-dipolar mechanism dominates the transition and overcomes the electric-quadrupolar one by over $12$ times for any rotation angle. Since in this case the rotation axis is parallel to the magnetic dipole moment, a purely magnetic dipolar transition should not show any modulation. This is indeed confirmed by the blue dotted line in Fig.~\ref{fig:stable}(c). The influence of the weaker quadrupolar channel is, however, manifested through interference, which modifies the full transition rate by up to $53.1\%$ [Fig.~\ref{fig:stable}(d)] either in the constructive or destructive way, depending on molecular orientation, leading to oscillations of significant amplitude. 

Due to the symmetry reasons, results for transition $2$ can be obtained by a simultaneous interchange of transition indices $1\leftrightarrow 2$ and rotations axes $y\leftrightarrow x$. Please note that if both states $1$ and $2$ were initially populated, the total emission rate would be an incoherent sum of the two corresponding contributions with weights determined by the population distribution. Since this would result in decreasing interference visibility, preparation of the molecule in only one of the states of the degenerate pair through selective pumping schemes would be beneficial. 

\section{Conclusions}
Combining results from quantum chemistry, nanooptics, molecular physics, and quantum mechanics, we have established a general framework to investigate the interplay of different multipolar transition channels with plasmonic nanoantennas. We have shown the tunability of the considered nanoantenna with respect to different multipolar contributions, that can be selectively enhanced through devoted illumination schemes. The nanoantenna was engineered to balance the rates corresponding to different multipolar transition mechanisms, unlocking the possibility of a considerable degree of interference. We have explored the possibility to superimpose and to coherently control the different emission pathways of quantum emitters originating from different multipolar contributions of the same transition. We have identified scenarios where the transition rate is enhanced but also suppressed below the free-space rate through the quantum mechanical interference of the different transition pathways. In such a scenario, a specific emitter brought in a specific position and orientation relative to the nanoantenna may remain excited through lifetimes enhanced with respect to the transition lifetimes of an isolated molecule. Further exploiting the possibility of complete suppression of a certain transition through interference is the key for many applications in the context of quantum computing, quantum storage, and quantum communication. 

\section{Methods}
\subsection{Field characteristics}
For numerical simulations of optical response of the nanoantenna and field distributions, the commercially available software package CST Microwave Studio\cite{CST} has been used. The software works with the finite-difference frequency domain method FDFD. Details of implementation of different multipolar emitters, calculations of fields and evaluation of the resulting transition rates can be found in the \textit{Supporting Information}.  

\subsection{Molecular characteristics}
Molecular characteristics rely on the TDDFT implemented in the TURBOMOLE software \cite{Furche2014, Armbruster2008, Kuhn2014, Weigend2010, Furche2000, Autschbach2011, Casida2009}.
The molecular structure of OsO$_3$ was optimized for the ground state with the functional BP86 \cite{Becke1988,Perdew1986} using def-SV(P) basis sets \cite{Schafer1992}, (BP86/def-SV(P)). Excitations were calculated at the BP86/def2-TZVPP level \cite{Eichkorn1997}, and - for checking the influence of the functional - additionally with the hybrid functional B3-LYP~\cite{Lee1988}. For estimating the influence of spin-orbit coupling also the two-component variant of TDDFT \cite{Armbruster2008, Kuhn2013, Kuhn2014} with respective bases, BP86/dhf-TZVP-2c \cite{Weigend2010} was used. The inclusion of spin-orbit coupling results in a small shift (by $0.02$ eV)  of the two excitation energies and in a negligible splitting (roughly by $0.001$ eV). The changes in transition moments $\mathbf{m}^{0\mathrm{j}}$ and $\mathbf{Q}^{0\mathrm{j}}$ are small, around $2\%$. This is much less than changes upon using a hybrid functional such as B3-LYP instead of a pure one (ca. $0.1$ eV for the energy and $10\%$ for the transition moments). Also the influence of small changes in the Os-O distance (by $2$ pm) on excitation energies, ca. $0.1$ eV, is much larger than that of the spin-orbit coupling. 

The output from the method includes a set of eigenstates characterized by transition energies from the ground state, as well as electric dipole, magnetic dipole, and electric quadrupole transition moments. 
The multipolar transition moments are calculated according to \cite{Furche2000,Autschbach2011}
\begin{eqnarray}
p^\mathrm{0j}_k &=& \sum_{i,a} \left( X^\mathrm{j}+Y^\mathrm{j}\right)_{i,a} \langle \phi_i|\hat{p}_k|\phi_a\rangle, \\
m^\mathrm{0j}_k &=& \sum_{i,a} \left( X^\mathrm{j}-Y^\mathrm{j}\right)_{i,a} \langle \phi_i|\hat{m}_k|\phi_a\rangle, \\
Q^\mathrm{0j}_{kl} &=& \sum_{i,a} \left( X^\mathrm{j}+Y^\mathrm{j}\right)_{i,a} \langle \phi_i|\hat{Q}_{kl}|\phi_a\rangle, 
\end{eqnarray}
where $|\phi_i\rangle$ ($|\phi_a\rangle$) are occupied (virtual) Kohn-Sham orbitals, respectively \cite{Kohn1999}, $X^\mathrm{j}$ and $Y^\mathrm{j}$ parametrize the transition density and can be calculated from the TDDFT response equation \cite{Casida2009}.
The multipolar moment operators in the SI system are defined as:
\begin{eqnarray}
\hat{\mathbf{p}}_k &=& -e\hat{r}_k,\\
\hat{\mathbf{m}}_k &=& \frac{ie\hbar}{2m_e}\left(\hat{\mathbf{r}}\times\nabla\right)_k,\\
\hat{\mathbf{Q}}_{kl} &=& -\frac{e}{2}\left(\hat{r}_k\hat{r}_l-\frac{1}{3}\delta_{kl}\hat{r}^2\right),
\end{eqnarray}
and the indices $k,l\in\{x,y,z\}$ enumerate components of vectors and tensors.

\begin{acknowledgement}
This is a pre-print of an article published in Nature Communications. The final authenticated version is available online at:
https://doi.org/10.1038/s41467-019-13748-4 
The authors acknowledge
Dr. Andrey Miroshnichenko for his advice on the implementation of the electric quadrupole. M.G. and C.R. acknowledge support by the Deutsche Forschungsgemeinschaft (DFG, German Research Foundation) – project number  378579271 – within project RO 3640/8-1.
K.S. \& P.G. acknowledge support from the 
Foundation for Polish Science (project no. Homing/2016-1/8)
within the European Regional Development Fund.
J.S. acknowledges support from the Karlsruhe
School of Optics and Photonics (KSOP).
The authors also thank the Deutscher Akademischer Austauschdienst
(PPP Poland) and the Ministry of Science and Higher Education in Poland.
\end{acknowledgement}

\begin{suppinfo}
 
\subsection{Implementation of multipolar sources}
Different molecular transition channels are modeled with different types of emitters implemented in the commercially available software package CST Microwave Studio (CST) operating with the finite-difference frequency-domain method \cite{CST}. The emitters are realized via oscillating electric currents in perfect electric conductors.
The model of the electric dipole emitter is based on the principle of a Hertzian dipole [Fig.~\ref{fig:source_implementation}(a,e)]: A finite current oscillates over a tiny length causing an electric dipole moment parallel to the direction of the oscillation. The magnetic dipole emitter consists of a circular, oscillating electric current resulting in an oscillating magnetic moment parallel to the symmetry axis and perpendicular to the oscillating current [Fig.~\ref{fig:source_implementation}(b,f)]. To embody the electric quadrupole, it is necessary to represent the corresponding electric quadrupole tensor. This can be done in two steps. First, the off-diagonal elements can be represented by the canonical electric quadrupole depiction of four alternating charges placed in the corners of a square [Fig.~\ref{fig:source_implementation}(c,g)]. In CST, this approach is realized via four oscillating currents pointing to/away from the imagined charges. This structure corresponds to a quadrupole tensor with $Q_{ij} = 0$ if $i = j$ and $Q_{ij} = Q_{ji}$. We refer to this emitter as the "off-diagonal electric quadrupole emitter" (off-
diag. EQ). Second, the diagonal elements can be accounted for by a linear arrangement of four charges (two of each kind), while the two charges in the center are of the same kind and closely spaced [Fig.~\ref{fig:source_implementation}(d,h)]. Two opposing, linearly aligned currents are used in CST to realize this case. This corresponds to a quadrupole tensor defined via the main diagonal with $Q_{ii} \neq 0$, and is later referred to as the "diagonal" electric quadrupole emitter (diag. EQ). The dimensions of the implemented emitters
are on a length scale of a few \AA{} and, therefore, are much smaller than the operational wavelength in the visible. This is an essential requirement, since the emitters must be operated far away from their intrinsic resonances.

\begin{figure}
 \includegraphics[width=8.6cm]{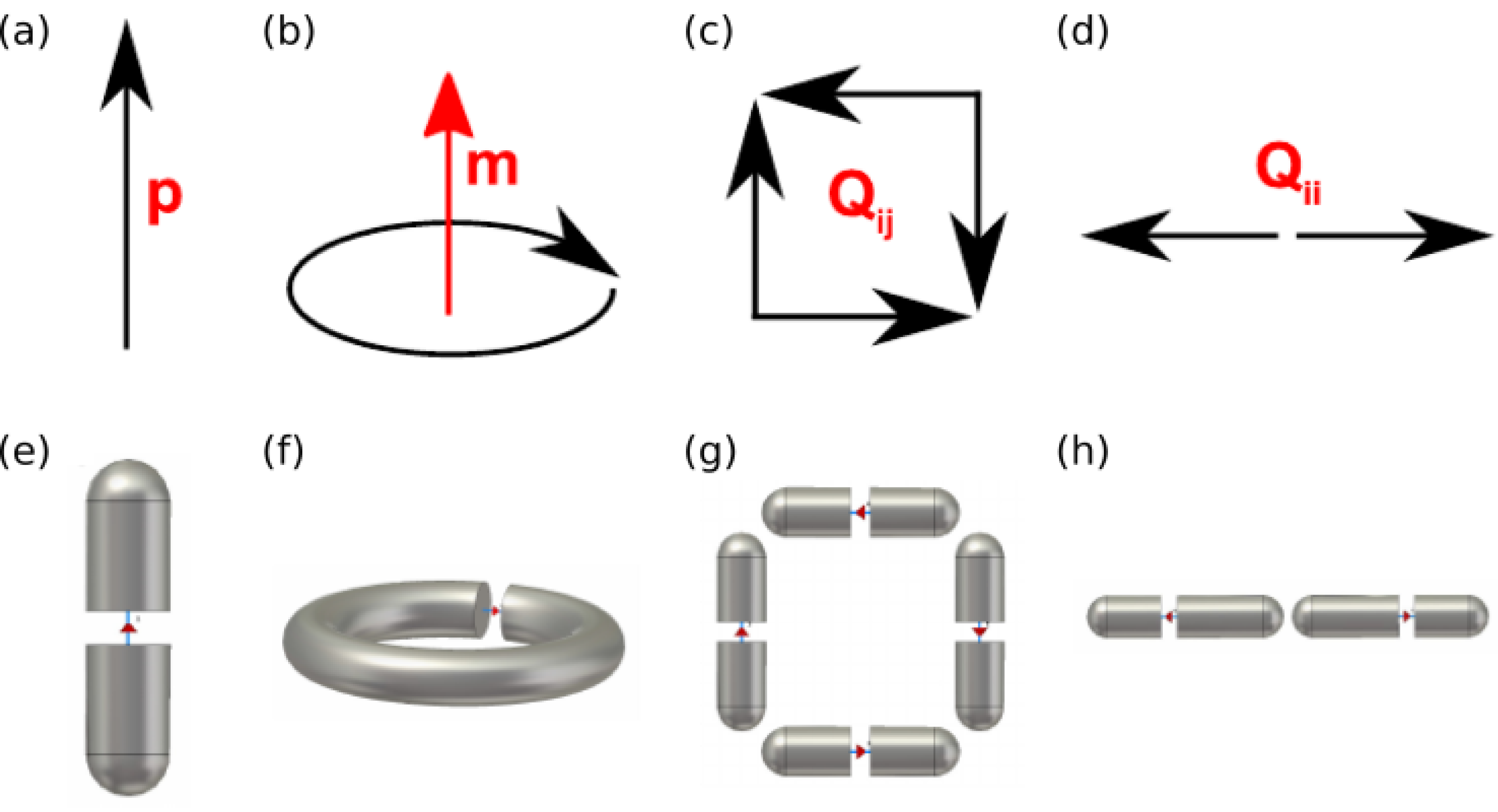}
  \caption{Top: The schematic representation of (a) the electric dipole, (b) magnetic dipole, (c) off-diagonal and (d) diagonal electric quadrupole emitters. The currents are displayed by black arrows. Bottom: (e-h) implementation of the emitters in CST Microwave Studio.}
  \label{fig:source_implementation}
\end{figure}

To test the quality of the emitters, we performed a multipolar expansion of the electric field $\mathbf{E}\left(\mathbf{r}\right)$
at the emission wavelength of the considered molecule at $\lambda = 553$ nm. For this, we use the
scattering coefficients $a_{nm}$ and $b_{nm}$ as defined in Ref.~\citenum{Muhlig2011}. 
The scattering coefficients are the multipole moments in spherical coordinates and can be transformed into Cartesian multipole moments for easier interpretation. The Cartesian electric and magnetic dipole moments ($n = 1$) are given by
\begin{equation}
\mathbf{p} = \left( \begin{array}{c} p_x \\ p_y \\ p_z \end{array} \right) = 
C_0 \left( \begin{array}{c} a_{11}-a_{1-1} \\ i(a_{11}+a_{1-1}) \\ -\sqrt{2}a_{10} \end{array} \right),
\mathbf{m} = \left( \begin{array}{c} m_x \\ m_y \\ m_z \end{array} \right) = 
cC_0 \left( \begin{array}{c} b_{11}-b_{1-1} \\ i(b_{11}+b_{1-1}) \\ -\sqrt{2}b_{10} \end{array} \right),
\end{equation}
with the constant $C_0 = i\sqrt{6\pi}\epsilon_0k^{-1}$, where $k$ is the wave number. 
The electric quadrupole moment (n = 2) can be calculated from
\begin{eqnarray}
\mathbf{Q} &=& \left( \begin{array}{ccc} 
Q_{xx} & Q_{xy} & Q_{xz} \\ 
Q_{yx} & Q_{yy} & Q_{yz} \\ 
Q_{zx} & Q_{zy} & Q_{zz} \end{array} \right) \\
&=& D_0 \left( \begin{array}{ccc} 
i(a_{22}+a_{2-2})-\frac{i\sqrt{6}}{2}a_{20} & a_{2-2}-a_{22} & i(a_{2-1}-a_{21})\\
a_{2-2}-a_{22} & -i(a_{22}+a_{2-2})-\frac{i\sqrt{6}}{2}a_{20} & a_{2-1}+a_{21}\\
i(a_{2-1}-a_{21}) & a_{2-1}+a_{21} & i\sqrt{6}a_{20}
\end{array} \right), \nonumber
\end{eqnarray}
with $D_0 = -\frac{i\sqrt{30\pi}\epsilon_0}{k^2}$. 
In Fig.~\ref{fig:source_check}
we plot the absolute square values of the scattering coefficients $|a_{nm}|^2$
for the electric emitters and $|b_{nm}|^2$ for the magnetic dipole emitter. We do not show the contribution of the magnetic multipolar coefficients $b_{nm}$ to the multipolar expansion of the electric emitters, since they are negligible, and analogously for the magnetic dipole emitter and the coefficients $a_{nm}$. We rescale the regarded quantities for the maximum value for all emitters in all orientations to be 1. 
An electric dipole emitter oriented in the $x$-direction has a dipole moment 
$p = p_x \sim a_{11} - a_{1-1}$.
This is confirmed by our results: in Fig.~\ref{fig:source_check}
, the only significant contributions to this emitter (ED x)
stem from $|a_{1-1}|^2$ and $|a_{11}|^2$. The other emitters in different orientations can be attributed to their corresponding Cartesian multipole moments, accordingly.

\begin{figure}
 \includegraphics[width=8.6cm]{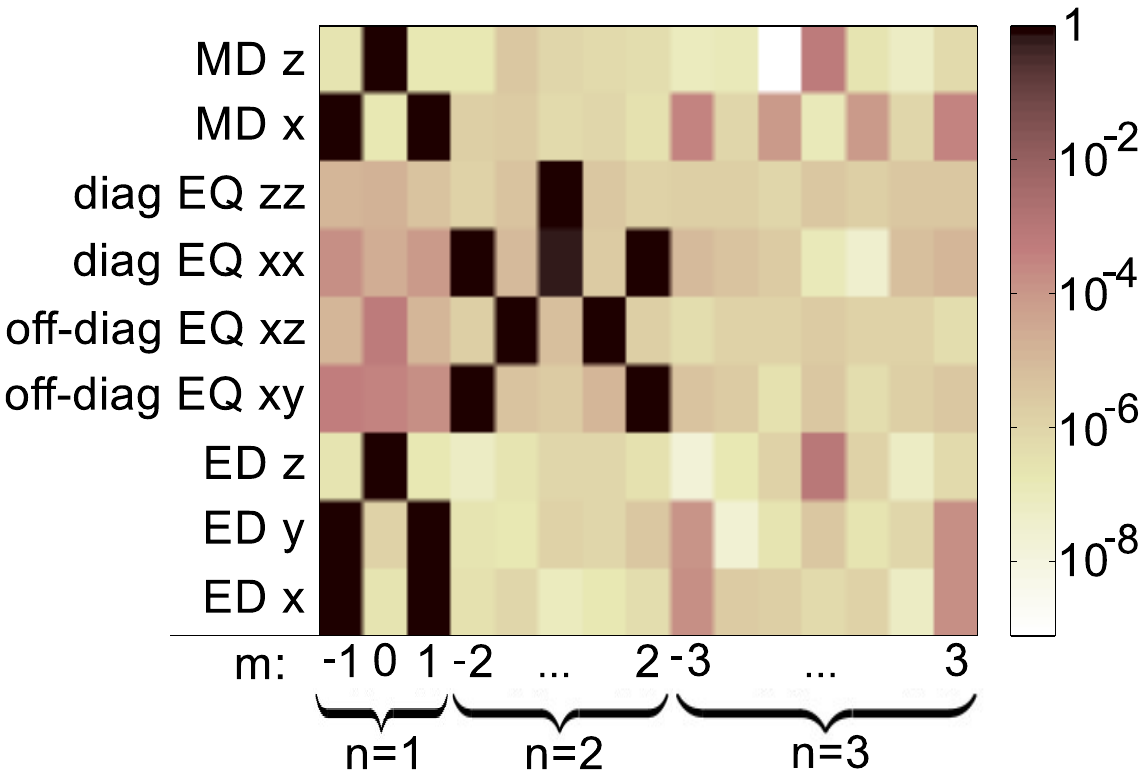}
  \caption{The absolute square values of the multipolar coefficients $|a_{nm}|^2$ for the electric emitters and $|b_{nm}|^2$ for the magnetic dipole emitter in different orientations at a free-space wavelength of $\lambda = 553$ nm. The lower case letters indicate the orientations of the emitters.}
  \label{fig:source_check}
\end{figure}

\subsection{Field distributions}
\begin{figure}
 \includegraphics[width=8.6cm]{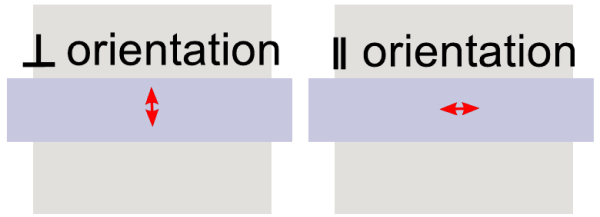}
  \caption{Two characteristic orientations
of the emitters relative to the patches of the nanoantenna correspond to a source, indicated by the red double-arrow, perpendicular or parallel to the patches. }
  \label{fig:patch}
\end{figure}
The transition rates $\Gamma$ are studied in function of the source orientation. For this purpose the total, vectorial, complex electromagnetic fields around the nanoantenna must be evaluated at the source position, which is a numerically demanding task. To reduce the complexity we make use of the fact that an arbitrarily oriented source can be decomposed into a superposition of corresponding sources of basic orientations. 
For dipole moments and for sources representing the diagonal components of the quadrupole tensor (see above) the basic orientations are (Fig.~\ref{fig:patch})
\begin{enumerate}
\item parallel to the patches along the $x$, or equivalently along the $y$ direction, 
\item perpendicular to the patches along the $z$ direction. 
\end{enumerate}
For sources that correspond to the off-diagonal elements of the quadrupole, the basic orientations are
\begin{enumerate}
\item parallel to the patches in the $xy$ plane aligned with the $x$ and $y$ axes, 
\item perpendicular to the patches equivalently in the $xz$ or $yz$ plane.
\end{enumerate}

For each type of source of normalized strength and of each basic orientation we calculate the distribution of both the electric and the magnetic fields around the nanoantenna, using an FDFD solver as discussed in section \textit{Implementation of multipolar sources} of this \textit{Supporting Information}. 
Consequently, the complex values of the spatial field components of the resulting fields are extracted. Since the multipolar sources are designed to act as point sources, but molecular transitions can hardly be pinpointed on sub-molecular dimensions, we replaced the field inside the innermost cubic $1$ nm$^{3}$ around the source by interpolated values. These interpolated values were generated by the Matlab internal function \texttt{interp3} based on the complete extracted fields in the cubic $1000$ nm$^3$ surrounding the source. Due to the field interpolation, the generated field values are not distorted by the actual shape of the different multipolar sources as shown in Fig.~\ref{fig:source_implementation} and potential numerical instabilities are also counteracted. The required field gradients are generated and processed in an analog fashion and finally the field values and gradients at the source position are calculated.

To model the composite response of such an intricate source representing a complicated actual molecular transition characterized by all its transition multipolar moments, we superpose the calculated complex electromagnetic field values and gradients with weights according to the ratio of the strengths of the corresponding transition multipolar moments. Due to the enhancement of the different transition channels through the nanoantenna and its fixed orientation, these superposition weights must be modified as the molecule is rotated with respect to the nanoantenna. This spatial-orientation-specific weight modification does not require completely new FDFD simulations, since solely the superposition of the sources of basic orientations must be adjusted accordingly. 
The thusly calculated fields and gradients enter Eqn.~1 for different rotation angles of the molecule around the different spatial axes by means of Eqn.~2 leading ultimately to the plots displayed in Fig.~3.

\subsection{Purcell enhancement}
Radiative and total Purcell enhancement factors calculated for different multipolar sources in basic orientations described above are shown respectively in Figs.~\ref{fig:Purcell_rad} and \ref{fig:Purcell_tot}. 

\begin{figure}
 \includegraphics[width=\textwidth]{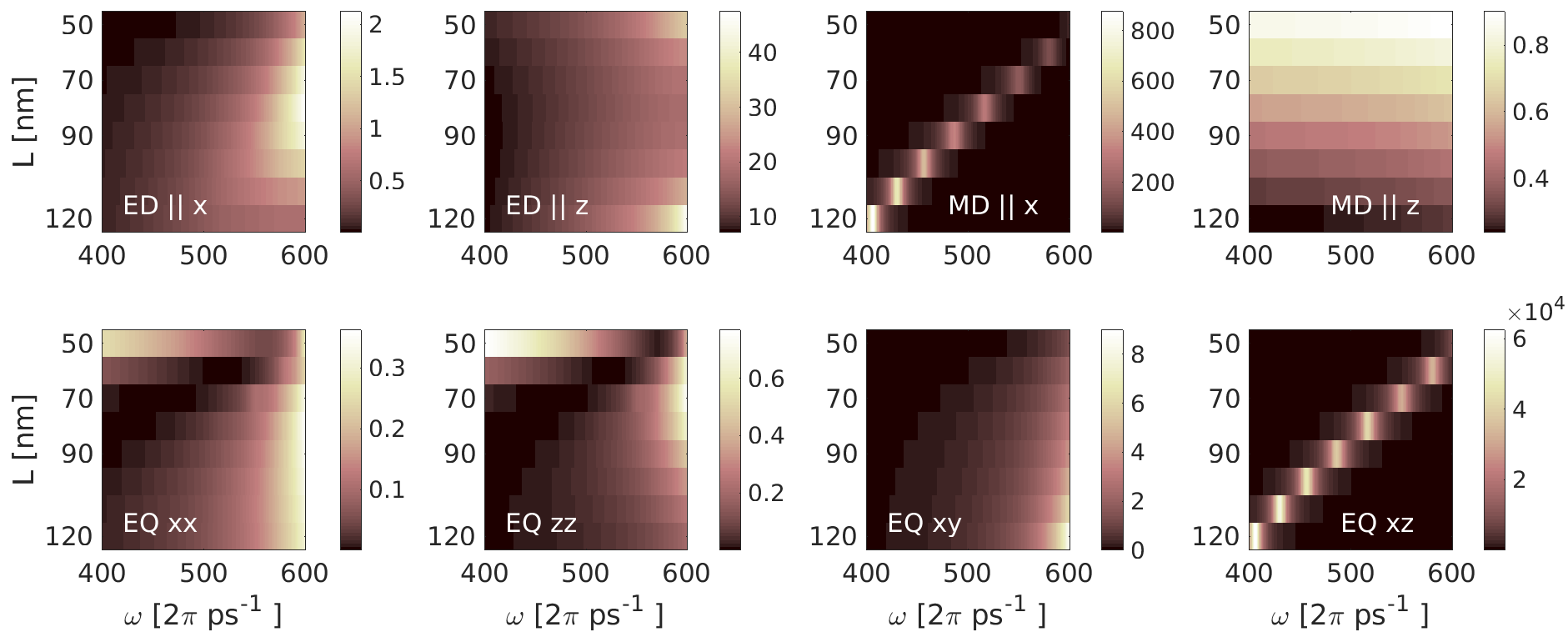}
  \caption{Purell enhancement of power radiated into far field by different multipolar sources (ED: electric dipole, MD: magnetic dipole, EQ: electric quadrupole) 
  in different basic orientations with respect to the nanoantenna discussed in this work. }
  \label{fig:Purcell_rad}
\end{figure}
\begin{figure}
 \includegraphics[width=\textwidth]{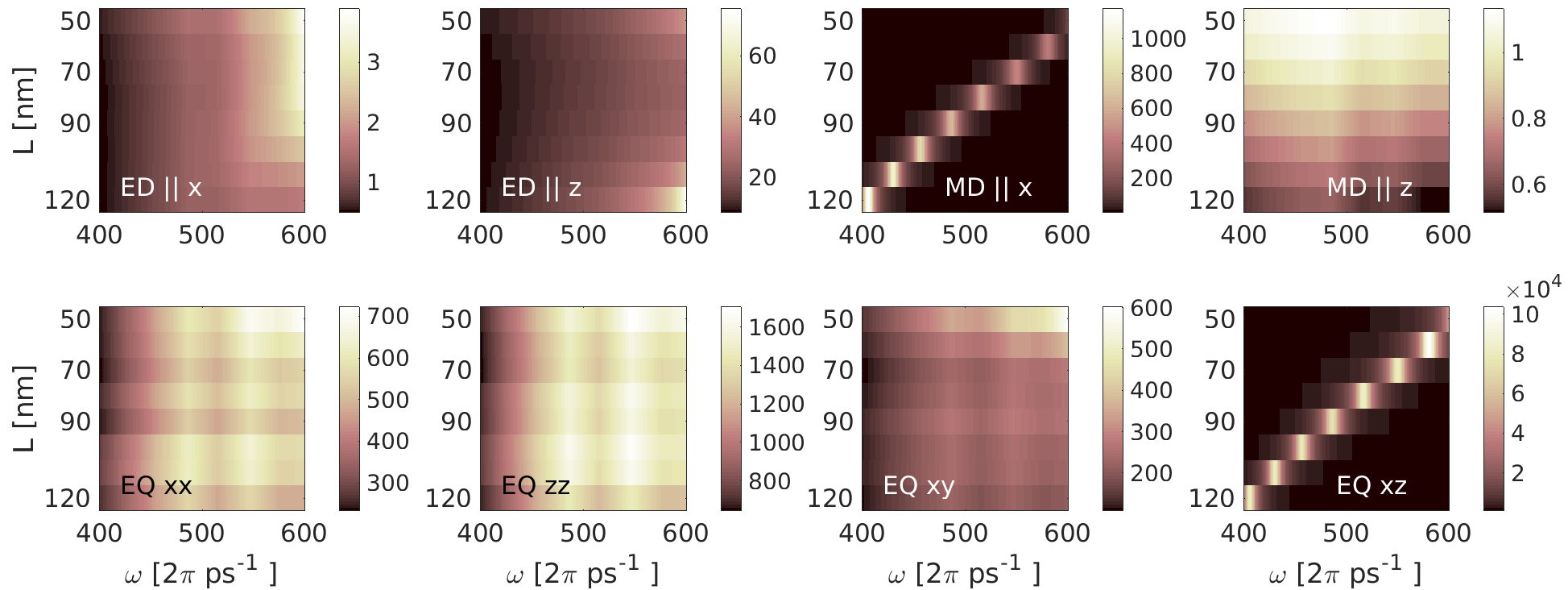}
  \caption{Purell enhancement of total power emitted by different multipolar sources (ED: electric dipole, MD: magnetic dipole, EQ: electric quadrupole) 
  in different basic orientations with respect to the nanoantenna discussed in this work. }
  \label{fig:Purcell_tot}
\end{figure}

\end{suppinfo}

\bibliography{hot.bib}

\end{document}